# Asymmetric Release Planning

## Compromising Satisfaction against Dissatisfaction

Maleknaz Nayebi, *Member, IEEE* and Guenther Ruhe, *Senior Member, IEEE*

**Abstract**—Maximizing satisfaction from offering features as part of the upcoming release(s) is different from minimizing dissatisfaction gained from not offering features. This asymmetric behavior has never been utilized for product release planning. We study *Asymmetric Release Planning* (ARP) by accommodating asymmetric feature evaluation. We formulated and solved ARP as a bi-criteria optimization problem. In its essence, it is the search for optimized trade-offs between maximum stakeholder satisfaction and minimum dissatisfaction. Different techniques including a continuous variant of Kano analysis are available to predict the impact on satisfaction and dissatisfaction with a product release from offering or not offering a feature. As a proof of concept, we validated the proposed solution approach called *Satisfaction-Dissatisfaction Optimizer* (SDO) via a real-world case study project. From running three replications with varying effort capacities, we demonstrate that SDO generates optimized trade-off solutions being (i) of a different value profile and different structure, (ii) superior to the application of random search and heuristics in terms of quality and completeness, and (iii) superior to the usage of manually generated solutions generated from managers of the case study company. A survey with 20 stakeholders evaluated the applicability and usefulness of the generated results.

---

# 1 INTRODUCTION

MAKING proper decision about the functionality of evolving software product releases is critical for the success or failure of a product. The process of understanding stakeholder needs and prioritizing them (see [1], [10], [59]) is a prerequisite for making good release decisions. Without a proper understanding of features and their value, product development becomes risky. The concept of satisfaction is largely related to *offering* features, while dissatisfaction is caused primarily by *not offering* them. However, there is an asymmetry between these two aspects: Some features generate satisfaction if delivered, but do not automatically create dissatisfaction if they are not offered. Vice versa, some features might create high dissatisfaction if not offered, but not necessarily create satisfaction if offered. We call this relationship between customer satisfaction and dissatisfaction an *asymmetric value impact* of a feature. The release planning process in consideration of asymmetric value impact of features is called *Asymmetric Release Planning* (ARP).

Release planning is defined as the process of selecting and assigning features to upcoming release(s) such that technological and effort constraints are satisfied and a stated utility function is maximized [62]. As part of this process, feature prioritization addresses the priority of a feature relative to the other ones. Release planning approaches for answering *what* and *when* to release problems have been modeled by defining planning objectives and constraints and by considering the feature values, feature dependencies, and stakeholder priorities.

We performed a literature analysis and found that 33 studies considered satisfaction or dissatisfaction as criteria for feature prioritization or subsequent release planning. However, only four of these studies jointly considered satisfaction and dissatisfaction (for further details, please see Section 7.1). In this paper, for the first time, we look at release planning in consideration of customers' satisfaction and dissatisfaction as competing criteria.

Customer[1] value or customer satisfaction were considered by several release planning approaches [46], [70], but none of them handled the conjoint effect of satisfaction and dissatisfaction on planning for future releases. This says, by considering either satisfaction or dissatisfaction as one planning objective, then the other objective should be considered as well. The asymmetric behavior of feature value requires a different modeling and solution approach when compared to the former (symmetric) planning methods because the results are expected to be different.

The main research questions addressed by this paper are:

**RQ1:** (Modeling) Which models exist for measuring satisfaction and dissatisfaction in the release management process?

**RQ2:** (Method) What release planning method creates trade-off solutions by accommodating the asymmetry between satisfaction and dissatisfaction?

**RQ3:** (Evaluation) For a real word case study, and for the method found in RQ2, how do the optimized plans (i) compare to plans generated randomly or by greedy search, (ii) are ranked by stakehold-ers, and (iii) compare to manual plans created by managers in terms of structure and quality?

---

• *M. Nayebi and G. Ruhe are with the Software Engineering Decision Support Laboratory, University of Calgary, Canada*  
*E-mail: {mnayebi, ruhe}@ucalgary.ca*

1. While formally a *customer* is a specific form of a *stakeholder*, both terms are used interchangeably in this paper to accommodate the terminology used in different contexts such as Kano (customer) analysis and stakeholder driven release planning.

## 2 MOTIVATING EXAMPLE

In this section, we motivate the use of ARP by a small sample project. For simplicity, we assume that we just have nine features F1 ... F9 and just one customer. All features have been evaluated by the customer in terms of satisfaction and dissatisfaction. Feature priorities are defined on a nine-point scale ranging from 1 (very low) to 9 (very high). For example, a satisfaction score of 9 for a feature F1 means that the customer is very satisfied if this feature is offered. However, this does not tell anything about the dissatisfaction of the customer in case the features are not delivered. Similarly, a dissatisfaction score of 9 for feature F9 means that the customer is very dissatisfied if this feature is not offered. Feature F7 has the same degree of dissatisfaction but differs in the degree of satisfaction once it would be offered. The resulted feature priorities are shown in Table 1.

The asymmetry is per feature. For example, F1 creates very high satisfaction if offered, but very low dissatisfaction if not offered. Offering F9 contributes to reducing customers dissatisfaction more than creating satisfaction among them. The asymmetry of F9 is reflected in Table 1 as the satisfaction for F9 is one while the dissatisfaction equals to nine.

To keep the example simple, features are assumed to consume an effort of a one person day for implementation. We further assume that the project has a capacity of three person days to implement features. Also, we only plan for the features of one (the next) release.

In total, there are $(9 \times 8 \times 7)/6 = 84$ possibilities to pick three features for the next release. Our proposed solution will determine six plans P1 ... P6. The plans are described in Table 2. Plan P1 provides the least satisfaction (= 6) but is best in the sense that it causes the least dissatisfaction (= 25). On the other end of the spectrum, P6 provides the highest satisfaction (= 27) but also causes the highest dissatisfaction (= 46). The values of the six plans are plotted in Figure 1.

In Table 2 the satisfaction of a plan is the sum of all individual feature satisfactions of that plan. Plan $P1 = \{F7, F8, F9\}$ creates a satisfaction of $3 + 2 + 1 = 6$. The dissatisfaction of $P1$ is the total dissatisfaction of all features not being offered, i.e., $1 + 2 + 3 + 4 + 7 + 8 = 25$.

We discuss three planning scenarios:

**Scenario 1**: Planning is directed towards maximizing satisfaction. This scenario would result in the

TABLE 2
Asymmetric release plans (to be) found by SDO.

| Plan ID | Feature sets | Satisfaction | Dissatisfaction |
|---|---|---|---|
| P1 | F7,F8,F9 | 6 | 25 |
| P2 | F5,F7,F8 | 12 | 27 |
| P3 | F5,F6,F7 | 14 | 28 |
| P4 | F3,F4,F5 | 24 | 38 |
| P5 | F2,F3,F5 | 25 | 40 |
| P6 | F1,F2,F3 | 27 | 46 |

    unique solution P6 of implementing feature set $\{F1, F2, F3\}$.

**Scenario 2**: Planning is directed towards minimizing dissatisfaction. This scenario would result in the unique solution P1 implementing feature set $\{F7, F8, F9\}$.

**Scenario 3**: Planning is based on both maximizing satisfaction and minimizing dissatisfaction (with equal weights). This scenario would result in two solutions P3 and P4 with feature sets $\{F5, F6, F7\}$ and $\{F3, F4, F5\}$, respectively.

We noticed that all solutions corresponding to the three scenarios are not only different in value, but also largely different in terms of the features offered. In summary, our proposed asymmetric release planning approach SDO will overcome former symmetric planning limitations in two main aspects:

(i) SDO considers the impact of both satisfaction and dissatisfaction that is caused by providing (resp. not providing) features.
(ii) SDO generates not just one solution, but a portfolio of trade-off solutions, compromising between satisfaction and dissatisfaction.

## 3 RQ1: MODELING FEATURE SATISFACTION AND DISSATISFACTION

The results of various studies showed that the relationship between satisfaction and dissatisfaction is asymmetric (see for example [60], [53]). So far, in software engineering, the conjoint relation between satisfaction and dissatisfaction has been ignored. Khurum et al. [32] provided a comprehensive value map for software products and reported that acquiring value for comprehensive feature prioritization needs careful analysis. Thus, software feature prioritization methods and, even further, release planning approaches have assumed that priorities are coming from stakeholders. So far, conjoint selection of satisfaction and dissatisfaction for prioritization or planning is optional. For example, in EVOLVE II, planning criteria could be *satisfaction* and *time to market*. However, a deeper analysis of users' satisfaction and dissatisfaction behavior showed that these two factors are two sides of the same coin.

TABLE 1
Satisfaction and dissatisfaction score of features.

| ID | Feature | Satisfaction score | Dissatisfaction score |
|---|---|---|---|
| F1 | Instant streaming | 9 | 1 |
| F2 | Multi-casting | 9 | 2 |
| F3 | Replay | 9 | 3 |
| F4 | Video on demand | 8 | 4 |
| F5 | Playlist | 7 | 7 |
| F6 | Video recommendation | 4 | 8 |
| F7 | Video history | 3 | 9 |
| F8 | Parental control | 2 | 9 |
| F9 | Share video | 1 | 9 |

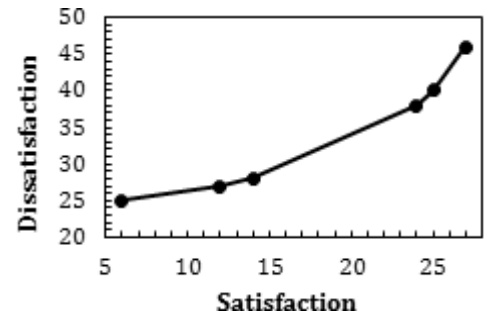

Fig. 1. Visualization of solutions listed in Table 3.

Selecting a model for measuring satisfaction and dissatisfaction is context specific. Hence, in what follows, we describe three models to predict the impact of offering or missing features in upcoming releases. The questions to be answered per feature and by all stakeholders are:

**One-point estimates:** The degree of satisfaction and dissatisfaction per feature and per stakeholder is the result of a one-point estimate.
**Pairwise comparison of features:** The degree of satisfaction and dissatisfaction per feature and per stakeholder is the result of aggregating pair-wise comparisons between features.
**Kano analysis:** For each feature a customer should answer functional and dysfunctional question. Satisfaction and dissatisfaction are calculated as the result of answering these two questions.

## 3.1 One-point estimates

Nine point evaluation of features have been used in software release planning and feature prioritization literature [62]. Applying it to ARP means that each feature is evaluated by all stakeholders twice to express her satisfaction of receiving a feature and her dissatisfaction of not receiving the same feature.

**Satisfaction:** To what extend you would satisfied by receiving this feature?
**Dissatisfaction:** To what extend you feel dissatisfied from not receiving this feature?

Because of the asymmetry between the two criteria, both questions have to be answered all the time. To express the satisfaction and dissatisfaction, each feature is evaluated by each stakeholder based on a nine-point scale defined below, where all even scores are described as the values between the two neighboured odd values.

1 - Very low
3 - Low
5 - Medium
7 - High
9 - Very high

For feature $F(n)$ and stakeholder $s$, the answers result in scores $sat(n,s)$ and $dissat(n,s)$. To differentiate between stakeholders importance, the weight of the stakeholder $wstake(s)$ is applied. It ranges from very low (= 1) to very high (= 9). A weight $wstake(s) = 0$ indicates that the stakeholder is not considered at all.

We compute weighted averages $S(n)$ and $DS(n)$ expressing the estimated total impact on satisfaction respectively dissatisfaction when delivering (not delivering) feature $F(n)$. The weighted averages are defined in Equations (1) and (2):

$$S(n) = \sum_s wstake(s) \times sat(n, s) / \sum_s wstake(s) \quad (1)$$

$$DS(n) = \sum_s wstake(s) \times dissat(n, s) / \sum_s wstake(s) \quad (2)$$

## 3.2 Pair-wise comparison of features by AHP

As an extension of the one-point estimates, applying the Analytical Hierarchy Process (AHP) [64] in this context means to perform pair-wise comparisons between all pairs of features from the perspective of both satisfaction and dissatisfaction. AHP allows decision-makers to assign ratio scale priorities to alternative features applying a sequence of pair-wise comparisons.

The process is applied independently for satisfaction and dissatisfaction perspective. Assuming two features called $F(n)$ and $F(m)$, a nine-point scale ranging from equally preferred (value = 1) over strongly preferred (= 5) to extremely preferred (= 9). A pair-wise comparison matrix $M$ is used to describe all these scores. For example, if feature $F(n)$ is strongly preferred to feature $F(m)$ in terms of satisfaction (for a given stakeholder), then $M(n, m) = 5$ and $M(m, n) = 1/5$. Applied for both objectives, AHP converts these evaluations to numerical values which are used as scores $S(n)$ resp. $DS(n)$.

Once these values are determined, weighted averages taking into account stakeholder weights $wstake(s)$ are determined similar to Equations (1) and (2). It is known that the whole AHP process provides more reliable scores. At the same time, it needs additional effort in comparison to one-point estimates.

## 3.3 Kano model

The traditional Kano questionnaire [30] overcomes point-wise evaluation by including two questions related to each feature. The first question (called *functional*) evaluates customers' reaction related to the presence of a specific feature. The second question (called *dysfunctional*) evaluates customers' reaction in the absence of the same feature:

**Functional question:** If feature F(n) is offered in the next release, how do you feel?
**Dysfunctional question:** If feature F(n) is not offered in the next release, how do you feel?

The answer to both questions has five choices as outlined below. In the traditional model, each customer must select exactly one single answer to each question [2]:

 (i) I like it that way
 (ii) It must be that way
 (iii) I am neutral
 (iv) I can live with it that way
 (v) I dislike it that way

---

2. We followed the order of questioning as proposed in the original paper. The impact of varying the order is considered a topic of future research.

TABLE 3
Kano evaluation table [30].

| Customer requirements | | Dysfunctional questions (D) | | | | |
|---|---|---|---|---|---|---|
| | | (i) Like | (ii) Must-be | (iii) Neutral | (iv) Live with | (v) Dislike |
| **Functional questions** (U) | (i) Like | Q | A | A | A | O |
| | (ii) Must-be | R | I | I | I | M |
| | (iii) Neutral | R | I | I | I | M |
| | (iv) Live with | R | I | I | I | M |
| | (v) Dislike | R | R | R | R | Q |

The Kano model categorizes features by combining the answers of the functional and dysfunctional questions [30]. As described in Table 3, each feature is classified into one of the following categories:

- Must be (*M*): The prerequisite features that the customer assumed as granted. Delivering these features do not affect the satisfaction level but prevent the occurrence of dissatisfaction among stakeholders.
- One-dimensional (*O*): The fulfillment of these features will linearly increment the degree of customer satisfaction.
- Attractive (*A*): These features are not explicitly requested. The absence of these features in the offered product will not cause dissatisfaction although the presence of them is expected to lead to greater customer satisfaction.
- Indifferent (*I*): The customer feels indifferent towards the availability or unavailability of these features. The feature might be considered as needed inside an organization, for example as an enabling technology.
- Reverse (*R*): Stated that the customer not only does not want this feature but even expected the feature not to be in the product. These features are avoided as they just waste resources without any desirable impact.
- Questionable (*Q*): This category is the indicator of a problem in phrasing the question or understanding the question. Therefore, these answers are not included in the analysis.

In what follows, we call these categories *Kano attributes*. In the traditional Kano model, each feature is assigned to exactly one of the above attributes. Based on this classification, Berger et al. [11] defined product satisfaction and dissatisfaction coefficients. They proposed to count the number of Attractive (#*A*) and the number of One-dimensional (#*O*) features and relate this number to the sum of #*A*, #*O*, #*M*, and #*I*. This also means that *R* and *Q* attributed features do not count in this evaluation. Stakeholders dissatisfaction with a product is defined analogously, with the focus on the number of Must-be features (#*M*) and a number of One-dimensional features (#*O*).

The two formulas for estimating stakeholders satisfaction and dissatisfaction of receiving a product are given as Equations (3) and (4) below:

$$Sat(P) = \frac{\#A + \#O}{\#A + \#O + \#M + \#I} \quad (3)$$

$$Dissat(P) = \frac{\#M + \#O}{\#A + \#O + \#M + \#I} \quad (4)$$

*Sat(P)* is a coefficient for the stakeholder satisfaction expected from receiving product *P* (compare [66]). Similarly, *Dissat(P)* is a coefficient telling how much providing product *P* avoids stakeholder dissatisfaction. From these equations:

- Increasing (#*A*) and (#*O*) will increase satisfaction. This aligns with reality, as the percentage of features contributing to satisfaction is increasing.
- By delivering more M and O features, we avoid customers dissatisfaction which occurs by not offering these features. In other words, increasing M and O will increase the amount of avoided dissatisfaction. This aligns with reality, as the percentage of features contributing to avoided dissatisfaction is increasing.
- Increasing (#*I*) will both reduce satisfaction and dissatisfaction. This aligns with reality, as resources are wasted for features not making direct contributions.

These coefficients were extracted experimentally. Later, other researchers analyzed these and provided more sophisticated formulations [43]. We relied on Berger's coefficient for simplicity as the baseline [11].

The *continuous Kano model* overcomes the limitations of the traditional model as it allows stakeholders to express their sentiments by selecting multiple responses for each question. These responses are aggregated with continuity in the feature scores (between categories) and include the degrees of importance per stakeholder. Extending the former fuzzy approach described by Lee et al. [34], continuous Kano analysis has four main steps:

**Step 1 – Normalizing stakeholders responses:** For a given feature and a specified stakeholder, $U_i$, $U_{ii}$, $U_{iii}$, $U_{iv}$, and $U_v$ [3] represent the normalized degrees of responses to the functional question. The different indexes refer to the five choices for each question. Normalization means that the sum of all five responses to a question is equal to 1. Similarly, $D_i$, $D_{ii}$, $D_{iii}$, $D_{iv}$, and $D_v$ represent stakeholders' response to the dysfunctional question.

**Step 2 – Calculating Kano attribute scores for each feature:** We denote the result of the evaluation of the attractiveness of feature *F(n)* in terms of Kano attribute *A* by stakeholder *s* as $score_A(n, s)$. As illustrated in Figure 2, this

---

3. To keep notation simple, we excluded any reference to both the feature and stakeholder in this context

score is determined by the summation over all the values that correspond to $A$ cells in Table 3.

$$score_A(n,s) = (U_i \times D_{ii}) + (U_i \times D_{iii}) + (U_i \times D_{iv}) \quad (5)$$

Equations (6) to (9) are defined following Lee et al. [34]. The equations are based on the classification of features given in Table 3. For example, $score_A(n, s)$ of Equation (7) is determined by summing up all parts of stakeholders answers resulting in category A (Attractive). We visualized the computation of Equation (7) in Figure 2.

The values $score_O(n,s)$, $score_M(n,s)$, and $score_I(n,s)$ are defined similarly. For One-dimensional, Must-be, and Indifferent Kano feature attributes, formulas are given as Equations (8), (9), and (10), respectively.

$$score_O(n,s) = (U_i \times D_v) \quad (6)$$

$$score_M(n,s) = (U_{ii} \times D_v) + (U_{iii} \times D_v) + (U_{iv} \times D_v) \quad (7)$$

$$score_I(n,s) = (U_{ii} \times D_{ii}) + (U_{ii} \times D_{iii}) + (U_{ii} \times D_{iv}) + (U_{iii} \times D_{ii}) + (U_{iii} \times D_{iii}) + (U_{iii} \times D_{iv}) + (U_{iv} \times D_{ii}) + (U_{iv} \times D_{iii}) + (U_{iv} \times D_{iv}) \quad (8)$$

**Step 3 – Calculating weighted averages for aggregating stakeholders score:** To differentiate between stakeholders importance, we use the weighted average for aggregating stakeholder scores as described in Equation (9).

To compute the overall attractiveness $F_A(n)$ of a feature $F(n)$, the individual scores $score_A(n, p)$ from all stakeholders are added to form the weighted average of Equation (7). The scores related to the other continuous Kano attributes are defined correspondingly.

$$F_A(n) = \frac{\sum_s (wstake(s) \times score_A(n,s))}{\sum_s wstake(s)} \quad (9)$$

**Step 4 – Calculating feature satisfaction and dissatisfaction values:** Modeling feature satisfaction and dissatisfaction follows the same idea as expressed for product evaluation by Berger et al. [11]. For this purpose, we extend the product related Equations (3) and (4) based on the number of (Boolean) occurrences to feature related (continuous) degrees of occurrences. This results in Equations (10) and (11). To calculate stakeholder satisfaction $S(n)$ of feature F(n), all stakeholder responses related to satisfaction elements (Attractive and One-dimensional) are divided by the sum of the Attractive, One-dimensional, Must-be and Indifferent portions of that feature:

$$S(n) = \frac{F_A(n) + F_O(n)}{F_A(n) + F_O(n) + F_I(n) + F_M(n)} \quad (10)$$

Similarly, $DS(n)$ is calculated by adding all responses with dissatisfaction elements (One-dimensional and Must-be) and dividing it by the total amount of relevant responses:

$$DS(n) = \frac{F_M(n) + F_O(n)}{F_A(n) + F_O(n) + F_I(n) + F_M(n)} \quad (11)$$

The detailed calculation of $S(n)$ and $DS(n)$ is illustrated in Appendix I.

## 4 RQ2: Asymmetric release planning

In this section we describe the objectives and constraints of asymmetric planning and introduce the solution approach SDO for providing trade off solutions to balance satisfaction and dissatisfaction.

### 4.1 Features

Decisions in product release planning are related to features and their assignment to releases. In the definition of features, we follow Wiegers and Beatty [77] who defined a product feature as *a set of logically related requirements that provide a capability to the user and satisfies the business objectives*. A more comprehensive list of release planning information needs is provided by Nayebi and Ruhe [47].

Let $F = \{F(1) \ldots F(N)\}$ be a set of $N$ candidate features for development during the upcoming $K$ product releases. A feature is called *postponed* if it is not offered in one of the next $K$ releases. Each release plan is characterized by a vector $x$ with $N$ components $x(n)(n = 1 \ldots N)$ defined as:

$$x(n) = k \quad \text{if feature } F(n) \text{ is offered at release } k \quad (12)$$

$$x(n) = K + 1 \quad \text{if feature } F(n) \text{ is postponed} \quad (13)$$

### 4.2 Objectives: Satisfaction versus dissatisfaction

The objective of our planning approach is to maximize stakeholder satisfaction and simultaneously minimize stakeholder dissatisfaction. These two objectives are independent and competing with each other. Pursuing each objective in isolation will create different release planning strategies.

For modeling of the satisfaction objective, we follow the proven concepts of the EVOLVE based algorithms, in particular, the most recent EVOLVE II [62]. For a given time horizon of $K$ releases, there is a discount factor making the delivery of a feature less satisfactory when it is offered later. While using a weighting (discounting) factor $w(k)$ for all

Fig. 2. Mapping between Equation (5) and Table 3.

releases $k = 1...K$, we assume that $w(K+1) = 0$, $w(1) = 1$ and

$$w(k) > w(k+1) \quad (k = 1 \ldots K-1) \quad (14)$$

This assumption implies that the value of delivering a feature will be the higher the earlier it is delivered. For a plan $x$ assigning features to releases, Total Satisfaction $TS(x)$ is defined in Equation (15). It is based on the summation of the discounted feature values $S(n)$ taken over all assigned features and all releases.

$$TS(x) = \sum_{k=1\ldots K} \sum_{n:x(n)=k} w(k) \times S(n) \to Max! \quad (15)$$

Total Dissatisfaction $TDS(x)$ of a plan $x$ follows the same idea as just introduced for satisfaction. The longer a feature is not offered, the higher the dissatisfaction. Similar to satisfaction, we introduce factors describing the relative degree of dissatisfaction between releases. $z(k)$ is the dissatisfaction discount factor related to release $k$. As dissatisfaction of non-delivery increases over releases, we assume $z(1) = 0$, $z(K+1) = 1$ and

$$z(k) < z(k+1) \quad (k = 1\ldots K) \quad (16)$$

If plan $x$ would not offer any features at all, total dissatisfaction $TDS(x)$ would be the summation of all feature dissatisfaction values. More general, if a feature is offered in release $k$, then this creates a dissatisfaction of $z(k) \times DS(n)$. If it is offered in the next release, no dissatisfaction is create at all. Total dissatisfaction $TDS(x)$ created by a plan $x$ is modeled as the summation of all adjusted feature values $DS(n)$, and this function needs to be minimized:

$$TDS(x) = \sum_{k=1\ldots K+1} \sum_{n:x(n)=k} z(k) \times DS(n) \to Min! \quad (17)$$

### 4.3 Resources

Implementation of features consumes effort. We make the simplifying assumption of just looking at the total amount of (estimated) effort needed per feature. The estimated effort for implementation of feature $F(n)(n = 1\ldots N)$ is denoted by $effort(n)$. When planning $K$ subsequent releases, the consumed effort per release is not allowed to exceed a given release capacity. For all releases $k (k = 1\ldots K)$, this capacity is denoted by $Cap(k)$.

More formally, a *feasible release plan* $x$ needs to satisfy all constraints of the form:

$$\sum_{n:x(n)=k} effort(n) \leq Cap(k) \text{ for k = 1. . . K} \quad (18)$$

We can also provide a more fine-grained model that subdivides the effort per features into more specific types of effort related to analysis, coding, and testing (as an example). The additional constraints resulting from that would not create principal new difficulties for the proposed solution approach but are ignored here to keep the model more simple.

### 4.4 ARP problem formulation

ARP extends the well-established symmetric release planning by considering both satisfaction and dissatisfaction and by treating them as independent criteria. This will be shown to have an impact on the number and structure of plans generated. Having $TS(x)$ and $TDS(x)$ as planning objectives, we are looking for trade-off solutions that maximizes satisfaction and minimized dissatisfaction.

Among all the plans fulfilling resource constraints (known as *feasible plans*), a plan $x^*$ is called a *trade-off solution* for ARP if no other plan exists that is better on one criterion and at the same time not worse in the other. This means that we are looking for feasible plans $x^*$ with the property that there is no other feasible plan $x^t$ (also called a *dominating plan*) such that:

(i) $TS(x^t) \geq TS(x^*)$
(ii) $TDS(x^t) \leq TDS(x^*)$, and
(iii) $(TS(x^t), TDS(x^t)) \neq (TS(x^*), TDS(x^*))$
(19)

**ARP Problem:** We consider a given set of features $F(n)$ with feature values $S(n)$ and $DS(n)(n = 1\ldots N)$. Among all the plans fulfilling resource constraints, the ARP problem is to find trade-off solutions for concurrently maximizing $TS(x)$ and minimizing $TDS(x)$. That means, ARP is the problem of finding trade-off release plans that are balancing satisfaction and dissatisfaction.

### 4.5 Solution Approach SDO

In its nature, the above ARP formulation is an (bi-objective) Integer Linear Programming (ILP) problem. ILP has been proven successful for solving the (symmetric) next release problem [73]. We propose an approach called *Satisfaction-Dissatisfaction Optimizer* or *SDO*. To address the bi-objective nature of the problem, we transform the integer bi-objective optimization problem ARP into a sequence of single objective problems for maximizing function $G(x, a)$ for varying α values:

$$G(x,a) = a \times TS(x) + (a-1) \times TDS(x)$$
among all feasible plans x and for all α from (0,1)
(20)

As the function $G(x, a)$ is to be maximized, the portion referring to minimize dissatisfaction is added with the factor $a - 1$. It is known from multi-criteria ILP [17] that:

- All solutions received from the parametric single-objective problem Equation (20) with parameter α varying between (0, 1), represent a trade-off solution for ARP.
- Different values of α may generate the same solutions for ARP. The ranges of α returning the same optimal solution are called *stability interval* of that solution.
- Iterative application of the parametric single-objective problem of Equation (20) cannot guarantee to determine the complete set of all non-dominated solutions (Pareto front).

In Section 3, we have outlined three methods for predicting the impact of offering or missing a feature. The results of each of these methods can serve as input for SDO. The methods are different in nature, in the effort needed and in the reliability of the predictions made. The one-point estimate is the simplest among the three methods. It was used for generating the input of the illustrative example in Section 2. With $N$ features and $S$ stakeholders, $2 * N * S$ estimates are required. The pair-wise comparison needs $N * (N-1)$ estimates and some subsequent eigenvalue computations. With its quadratic effort, this method is applicable only for low to a mid-size number of features. Compared to one-point estimates, its benefit is that evaluations are expected to be more reliable, as they are based on comparisons between all pairs of features.

Kano analysis is the most advanced among the three methods. The continuous method requires $10 * N * S$ evaluations. Per feature and per stakeholder, 100 points can be allocated for the five possible answers. The computation of the S(n) and DS(n) scores is straight forward as discussed in Section 3.

## 4.6 SDO tool

We are proposing a method of solving a sequence of single-criterion optimization problems, each of them generating a new or an existing trade-off solution. The step-size for varying the parameter can be selected by the concrete problem. For the implementation of SDO, we apply the commercial optimizer Gurobi [27] version 6.4 and its interface MATLAB to manage data. We call this the *SDO tool*. Gurobi is a set of optimization libraries for linear programming, quadratic programming, and mixed-integer programming. We used the free academic license available for Gurobi to develop our application. For problems up to several hundreds of features, the solution gained from running single-objective optimization are proven to be 100% optimal. The optimality status is displayed by the Gurobi optimizer.

While we rely on Gurobi, the results of our approach in principle do not depend on the underlying optimizer. Depending on the solver, there might be differences in the scalability of the approach, i.e., the size of problems we are able to solve. We measured the time required to determine the optimized plans when using Gurobi on an Intel(R) core i7-2620M CPU@ 2.70 GHZ computer. On average, SDO computed a Pareto solution (one iteration of the approach) in 54 ms which seems to be sufficiently fast from an application perspective.

The SDO implementation (code snippets in MATLAB) is available as a complementary material for this paper, and the environment setup and overview of the implementation are described in Appendix II. The step size (α) is a parameter in the SDO tool that needs to be defined for each problem. Defining it depends on the specific problem context and the targeted completeness of the solution set.

## 5 RQ3: Evaluation

To prove the applicability and usefulness of SDO and the quality of its results, we performed a case study [63] with a company developing mobile apps. For feature value prediction, we are using the Kano model.

Analysis of the development and usage of mobile apps is an emerging area of research. Properly addressing customer concerns is critical in the highly competitive and dynamic environment. In the context of mobile app stores, we study the ARP process to balance stakeholders' satisfaction and dissatisfaction for proposing (new) mobile apps [45] [48]. For that purpose, we looked at Over-The-Top (OTT) TV services [29] apps offered in the Android app store market and plan for the next release (K = 1) of that product.

Following the guidelines of Runeson and Hoest [63] for conducting case studies, we describe the design, the data collection process, the analysis of data, and the reporting of results for our case study. In RQ3, we evaluate different aspects of improvement from using SDO in comparison to plans generated from random and heuristic search and plans generated from human experts.

### 5.1 Data collection and preparation

Our data collection and preparation was organized towards elicitation of features, effort estimation, and feature evaluation by stakeholders. We describe the steps and their results in more detail:

**Stakeholders:** The case study company reached out to a large group of direct stakeholders (potential users of the app). Since we did not have access to their data, we invited 24 software engineering graduate students to serve as stakeholders. Even though students were not a direct customer of the company, they were familiar with the domain (OTT services) and were considered to be representative for the purpose of this case study.

**Weight of stakeholders:** The survey participants provided a self-evaluation in terms of their familiarity with OTT services and mobile applications. At the beginning of the survey, stakeholders stated their domain expertise on a Likert scale ranging from one to nine. We used this value as the weight of stakeholders for the planning process.

**Features**: The pool of candidate app features was extracted from the description of 261 apps, all of them providing media content over the Internet without the involvement of an operator in the control or distribution of the content (OTT service). A commercial text analysis tool was used to retrieve 42 candidate features. Domain experts evaluated the meaningfulness of extracted features and eliminated the phrases which did not point into any OTT feature. Feature extraction itself was managed by the case study company and resulted in 36 features further investigated.

**Feature values:** To predict the impact of offering versus missing features, we applied the Kano analysis outlined in Section 3. We performed a survey with a continuous Kano design and asked the two types of questions (functional and dysfunctional) that were introduced in Section 3.3. For each feature, each of the stakeholders expressed the percentages that the feature matches one of the five possible answers per question.

**Effort:** The effort for developing each feature was estimated by domain experts within the company. A product manager and two senior developers estimated the effort needed (in person hours) to develop each feature. They applied a triangular (three point) effort estimation to estimate the optimistic, pessimistic and most-likely effort amount needed to deliver a feature. The three estimates were combined using a weighted average with weighting factors of one, one and four, respectively. This weighting technique is originated from the Program Evaluation and Review Technique (PERT) [39].

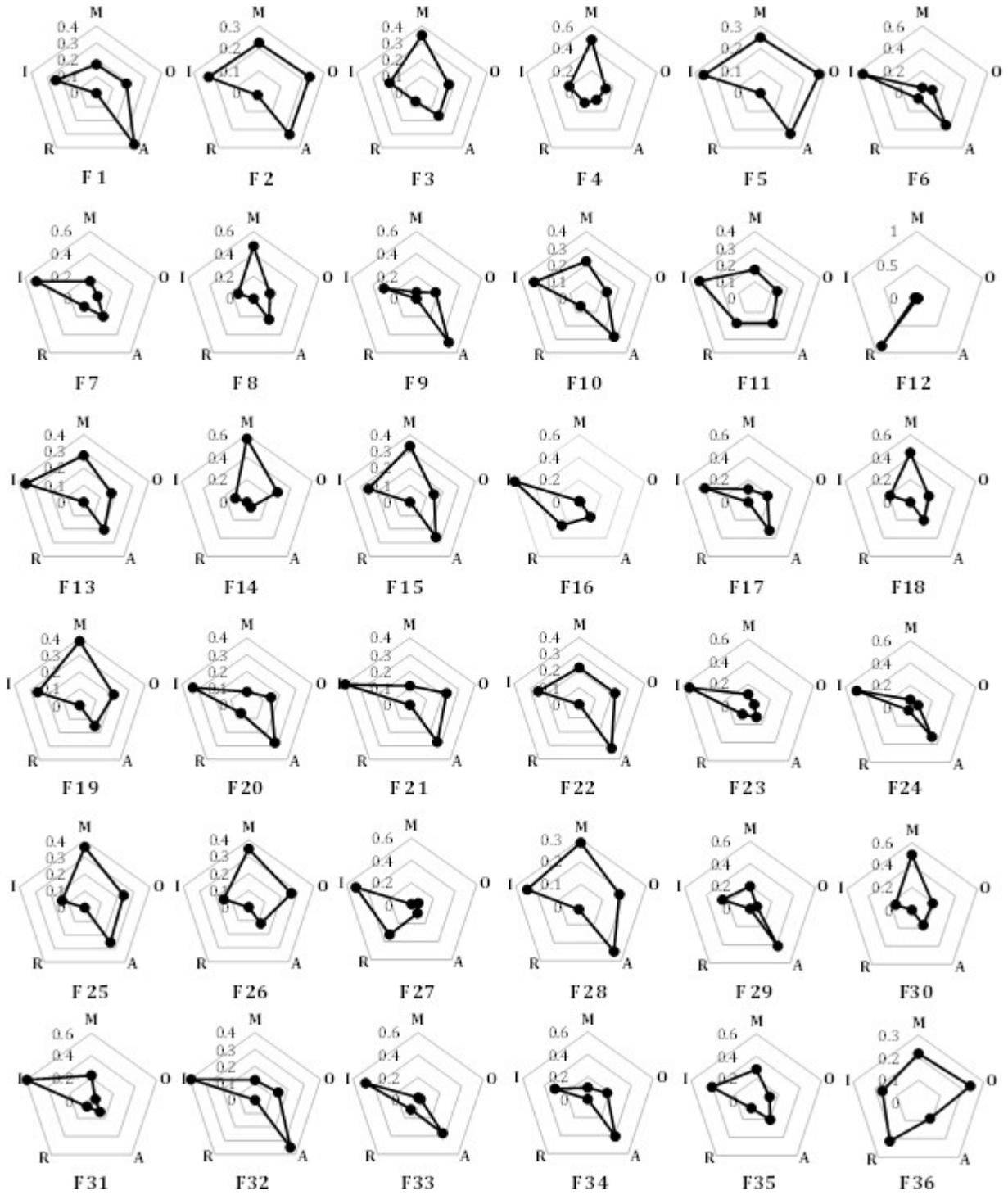

Fig. 3. Continuous Kano evaluation of the 36 case study features. The values in each of the five categories represent the degree of membership of features in the $M$, $O$, $A$, $R$, and $I$ category.

**Capacity**: To show the behavior of SDO under different scenarios of tight, medium, and more relaxed resource availability, we ran three concurrent scenarios with release capacities Cap(1) of 112.7 (lower bound), 367.4 (most probable), and 625.5 (upper bound) person hours, respectively.

Lists of features, effort estimation, continuous Kano survey, and results are available online[4]. For computing the

---

4. http://www.ucalgary.ca/mnayebi/tools-and-data-sets

Perato solutions, we varied α and found that increasing the granularity of the step size beyond 0.001 does not provide additional insight.

### 5.2 Feature satisfaction and dissatisfaction analysis

Using the data introduced above, we calculated stakeholder scores per features using Equations (5) to (8) and aggregated the scores per feature across stakeholders as shown in Equation (9). The resulting Kano attributes are presented as spider charts in Figure 3 for all 36 features. We observe that features mostly have aspects of all characteristics, but the difference is on the degree of emphasis for them. For example, F1 has mainly considered an attractive feature, contributing toward satisfaction. Similarly, F14 is mainly seen as a must-be feature. Thus, not offering it, would create substantial dissatisfaction. In Appendix I, the detailed calculation of Kano attributes is provided for sample feature F(15).

For better comparison across all features, we created another spider chart. In Figure 4, the red (dark grey) and (light) grey areas reflect the aggregated stakeholder dissatisfaction and satisfaction scores per feature, respectively. In this figure, the asymmetry between satisfaction and dissatisfaction for different features is reflected. Following clockwise ordering, features are arranged in decreasing level of satisfaction. Once the dissatisfaction surface extends the limits of the satisfaction (gray) surface, this indicates that the corresponding feature has stronger *Must-have* and *One-dimensional* attributes. Consequently, these features are contributing more toward avoiding customers' dissatisfaction rather than to satisfaction.

### 5.3 Analysis of solutions generated from SDO

For the purpose of comparison and evaluation, we applied SDO for planning 36 features considering the described constraints and resources. We first look into the portfolio of features generated by SDO. For the three scenarios (different levels of capacity) we study the structure and diversity of them.

Having all the data needed, we now analyze the results from applying SDO to solve the case study problem. This is done for the three varying capacity levels. A set of alternative solutions was generated by SDO for each level. In total, over the three capacity levels, 14 trade-off solutions (plans) were generated. All these release plans are presented in Table 4. Each line in the table represents one plan composed of the features offered, along with values of the objective functions (satisfaction and dissatisfaction) and the effort that is needed to implement the plan. For simplicity, we use *Fn* instead of *F(n)* to refer to features.

Each plan represents one possible way to balance between satisfaction and dissatisfaction of stakeholders. Based on the equivalence between parametric and multi-objective optimization [17], the solutions received from automatically running a sequence of single-criterion problems Equation (20) represent Pareto solutions for ARP.

### 5.4 Comparison with random and heuristic search

As formulated in Objective 2, we are interested in the comparison between the quality of SDO solutions in comparison

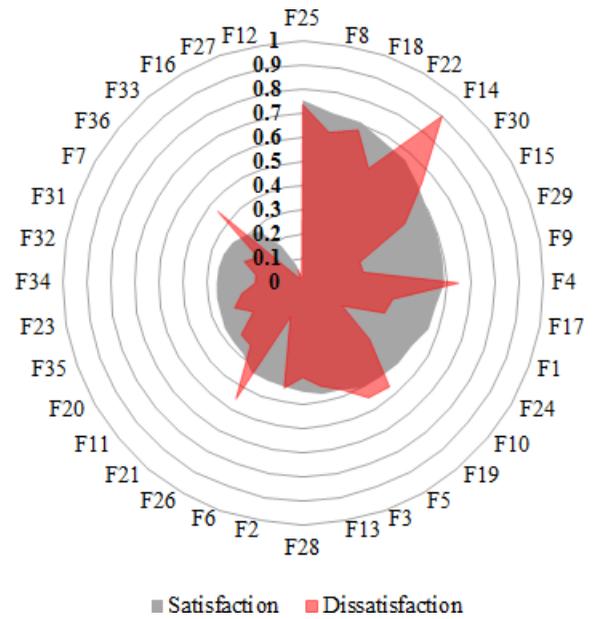

Fig. 4. Satisfaction - Dissatisfaction profile of features F(n) arranged in a clockwise order according to decreasing level of satisfaction.

to those obtained from (i) random search and (ii) heuristic search.

#### 5.4.1 Comparison with random search

Using random search as a baseline for comparison has been introduced by Acuri and Briand [3]. It has been applied in various contexts since then, see for example [80]. We used it here for the same purpose as a straw man. Random search is selecting a feature randomly as long as the effort for implementing that feature is less than the available capacity. The procedure is summarized as Algorithm 1. The results of running 1,000 replications for the case of medium capacity are shown in Figure 5-(a). The results showed that SDO generated solutions strongly dominate all the 1,000 solutions generated by random searching. The best random solution is still outperformed by 37% by one of the SDO solutions (having 12% better satisfaction and 25% better dissatisfaction performance).

#### 5.4.2 Comparison with heuristic search

As ARP is in the class of NP-complete problems [24], the question of finding light-weight heuristics arises. Greedy heuristics are widely used to solve combinatorial problems and are used in practice to easily create release plans [62]. The general greedy principle is to select the best local features at each iteration, where the definition of locally best varies between the heuristics [18]. With no backtracking, greedy solutions are fast and often "good enough".

**Algorithm 1** Random selection of features

Set Capacity, Remaining_capacity;
**do**
Select a random feature between F1 to FN;
Remaining_capacity = Capacity - Effort(n)
**while** No Feature with effort <= Remaining_capacity

TABLE 4
Results of asymmetric planning: Feature plans, satisfaction and dissatisfaction levels for each plan, stability of results and effort needed to implement optimized plans for three levels of capacity.

| Capacity = 112.7 | | | |
|---|---|---|---|
| Plan | Satisfaction | Dissatisfaction | Effort |
| F1 F3 F5 F14 F21 F22 F25 | 4.713 | 8.974 | 111.4 |
| F1 F3 F5 F21 F22 F25 F30 F32 | 5.031 | 9.030 | 111.5 |
| F1 F3 F5 F9 F18 F22 F25 F30 F32 | 5.236 | 9.112 | 112.4 |

| Capacity = 367.4 | | | | |
|---|---|---|---|---|
| Plan | Plan | Satisfaction | Dissatisfaction | Effort |
| Plan 1 | F1 F2 F3 F4 F5 F9 F11 F14 F15 F17 F18 F19 F21 F22 F23 F25 F30 F32 F36 | 10.024 | 4.301 | 360.2 |
| Plan 2 | F1 F2 F3 F4 F5 F9 F11 F14 F15 F17 F18 F19 F21 F22 F24 F25 F30 F32 F36 | 10.184 | 4.369 | 366.9 |
| Plan 3 | F1 F2 F3 F4 F5 F9 F10 F11 F14 F15 F17 F18 F19 F21 F22 F24 F25 F30 F32 | 10.394 | 4.469 | 362 |
| Plan 4 | F1 F2 F3 F5 F6 F9 F10 F11 F14 F15 F17 F18 F19 F21 F22 F23 F24 F25 F30 F32 | 10.597 | 4.710 | 366.3 |

| Capacity = 625.5 | | | |
|---|---|---|---|
| Plan | Satisfaction | Dissatisfaction | Effort |
| F1 F2 F3 F4 F5 F7 F8 F9 F10 F11 F13 F14 F15 F17 F18 F19 F21 F22 F23 F25 F26 F28 F30 F32 F35 F36 | 13.298 | 1.413 | 616 |
| F1 F2 F3 F4 F5 F8 F9 F10 F11 F13 F14 F15 F17 F18 F19 F21 F22 F23 F24 F25 F26 F28 F30 F32 F35 F36 | 13.485 | 1.439 | 618.2 |
| F1 F2 F3 F4 F5 F6 F7 F8 F9 F10 F11 F14 F15 F17 F18 F19 F21 F22 F23 F24 F25 F26 F28 F30 F31 F32 F36 | 13.784 | 1.508 | 622.5 |
| F1 F2 F3 F4 F5 F6 F7 F8 F9 F10 F11 F13 F14 F15 F17 F18 F19 F21 F22 F23 F24 F25 F26 F30 F32 F35 F36 | 13.786 | 1.509 | 625.3 |
| F1 F2 F3 F4 F5 F6 F7 F8 F9 F10 F11 F13 F14 F15 F17 F18 F19 F21 F22 F23 F24 F25 F28 F30 F32 F35 F36 | 13.825 | 1.629 | 623.2 |
| F1 F2 F3 F4 F5 F6 F7 F8 F9 F10 F11 F13 F14 F15 F17 F18 F19 F21 F22 F23 F24 F25 F28 F30 F31 F32 F35 | 13.866 | 1.833 | 624.6 |
| F1 F2 F3 F4 F5 F6 F7 F8 F9 F10 F11 F13 F14 F15 F17 F18 F19 F20 F21 F22 F23 F24 F25 F28 F30 F31 F32 | 13.876 | 1.896 | 623.6 |

TABLE 5
Comparison of results between SDO generated plans and the results of eight heuristics for three different effort levels.

| ID | Algorithm | FACTOR | Marker | Dominated by SDO plan | Identical to SDO plan | New Pareto plan |
|---|---|---|---|---|---|---|
| H1 | Algorithm 2 | Satisfaction | ∗ | 0 | 3 | 0 |
| H2 | Algorithm 2 | Dissatisfaction | ■ | 2 | 1 | 0 |
| H3 | Algorithm 2 | Satisfaction/Effort | ♦ | 2 | 1 | 0 |
| H4 | Algorithm 2 | Dissatisfaction /Effort | - | 1 | 2 | 0 |
| H5 | Algorithm 2 | Satisfaction + Dissatisfaction | | 3 | 0 | 0 |
| H6 | Algorithm 3 | (Satisfaction + Dissatisfaction)/Effort | ○ | 2 | 0 | 1 |
| H7 | Algorithm 3 | Satisfaction, Dissatisfaction | | 3 | 0 | 0 |
| H8 | Algorithm 3 | Satisfaction/Effort, Dissatisfaction/Effort | × | 3 | 0 | 0 |

The quality of the solutions often depend on the problem structure and the instance of the problem. It can be quite far from the optimum in specific instances.

We applied the search for greedy solutions to ARP. In total, we defined and compared eight greedy heuristic approaches. Therein, *locally best* selection for the next feature is defined related to satisfaction, dissatisfaction or a combination of both. As another variation point, definition of *locally best* was done with and without consideration of effort consumed per feature.

All heuristics are instantiations of either Algorithm 2 or Algorithm 3. For Algorithm 2, we selected just one factor for ranking the features (referred to as FACTOR in the algorithm).

The factors used in the greedy heuristics are as follows:

- Satisfaction,
- Dissatisfaction
- Satisfaction/effort
- Dissatisfaction/effort
- Satisfaction + dissatisfaction
- (Satisfaction + dissatisfaction)/effort.

We used Algorithm 3 to define greedy solutions based on the application of two concurrent ranking criteria, applied alternatively between iterations. This is called *Two factors greedy heuristic* and it uses the following pairs of factors:

- [Satisfaction, dissatisfaction], and
- [Satisfaction/effort, dissatisfaction/effort].

For example, using Algorithm 2 with [Satisfaction, dissatisfaction] factors, the first feature picked is the best in terms of satisfaction. The second feature selected is the one creating the highest dissatisfaction if it is not selected.

---

**Algorithm 2** One factor greedy heuristic

Rank features in decreasing order of FACTOR;
**for** $n$ = 1 to $N$ **do**
  Select f(n) where f(n) is ranked higher than other features and was not selected before;
  **if** (Effort(n) < Capacity - Total Effort) **then**
    Total Effort + = Effort (f(n));
  **end if**
**end for**
**While**(Total Effort <= Capacity)

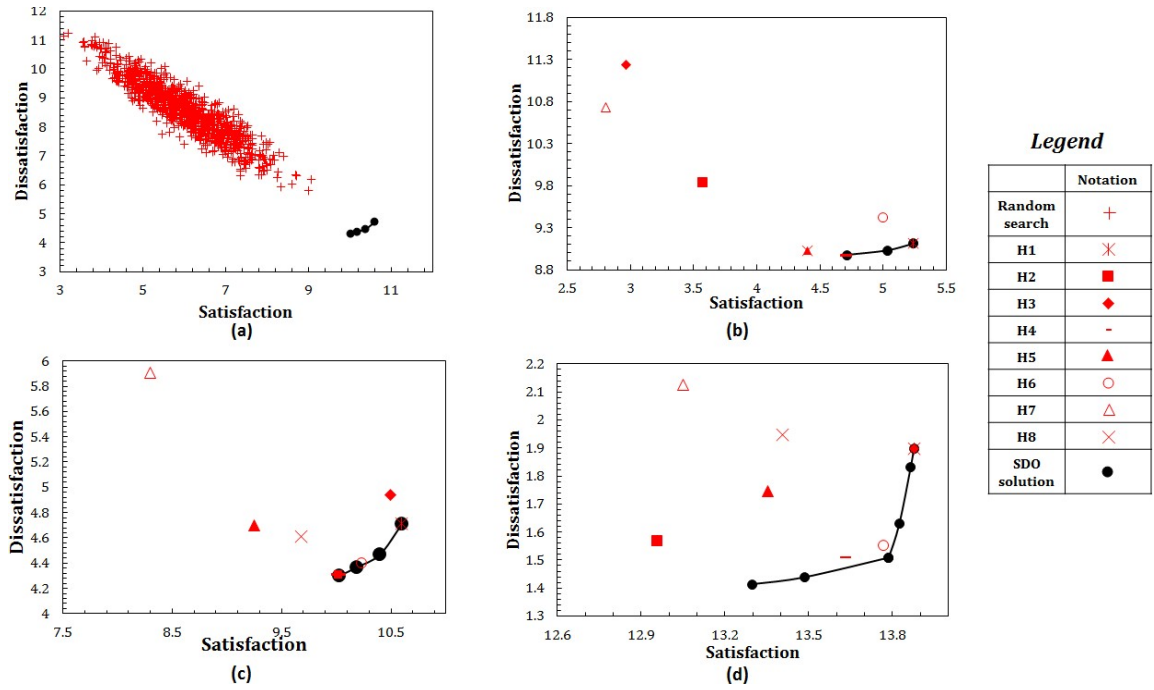

Fig. 5. Comparison between results of SDO generated plans with (a) a random search (capacity level 112.7) and the results of eight heuristics for three different capacity levels (b) 112.7, (c) 367.4 and (d) 625.5.

Now, the criteria are changed again, and the second best feature in terms of satisfaction is picked. The algorithm terminates when no feature can be added because of the capacity constraint. In Figure 5 (b) to (d), we summarize the results of running different heuristic factors in Algorithm 2 and Algorithm 3.

From running SDO for three levels of capacity, we received 14 Pareto solutions. By running eight different heuristics, we got in total 24 solutions. 66.7% of the heuristic plans were dominated with at least one of the SDO solutions. Also, one new Pareto solution was retrieved by heuristic H6 which was not found by SDO (the rest were identical to existing SDO solutions). We also noticed that heuristic H2, which is aimed to optimize towards the dissatisfaction criterion, does not even achieve a solution which is best related to that criterion. This demonstrates that heuristics are fast and conceptually easy, but often not good enough. The comprehensiveness of solutions generated from SDO in conjunction with their guaranteed quality is considered a strong argument in favor of the proposed SDO approach.

Comparison between solutions generated from SDO against the eight different heuristics showed that:

- Using stakeholders satisfaction in Algorithm 2 (H1 in Table 5) performed best compared to other heuristics, as it found three solutions identical to the Pareto solutions found by SDO.
- Using dissatisfaction/effort (H4 in Table 5) also found two solutions which were identical to one of the SDO solutions. One of the solutions found by this heuristic was dominated by SDO plans.
- Among all the heuristics, using Algorithm 3 (H7 and H8 in Table 5) had the worst performance. Both heuristics found solutions to ARP which were all dominated by SDO.

The comparison of solutions received from SDO and from the different heuristics are presented in Table 5.

### 5.5 Expert evaluation of plans

Besides comparing the results of SDO with other algorithms, we have performed an external evaluation from running two rounds of surveys with actual stakeholders to make sure about the relevance of the analysis with managers' need [49]. Industrial evaluation of release planning models is known to be inherently difficult. The results of a systematic literature review by Svahnberg et al. [70] showed a deficit in the industrial evaluation of release planning methods. This was confirmed by a recent survey of software release planning models conducted by Ameller et al. [2]. The authors found that since 2009, only four release planning papers were published that include at least one co-author from industry.

We conducted a survey to understand stakeholders preference among the various plans generated. The survey

---

**Algorithm 3** Two factor greedy heuristic

Arr A[] = Rank features in FACTOR decreasing order; Arr B[] = Rank features in FACTOR' decreasing order; Arr C[] = Merge array A[] with array B[]
**for** $n$ = 1 to $N$ **do**
  **do**
    Select f(n) where f(n) is ranked higher than other features in C[] and was not selected before;
    **if** (Effort(n) < Capacity - Total_Effort) **then**
      Total_Effort + = Effort (f(n));
    **end if**
  **while** (Total_Effort < Capacity)
**end for**

TABLE 6
Stakeholder ranking of plans for the four plans of Table 5 calculated for Capacity = 367.4. All plans were evaluated from satisfaction perspective.

| Stakeholder | Plan 1 | Plan 2 | Plan 3 | Plan 4 |
|---|---|---|---|---|
| Student 1 | 4 | 1 | 2 | 3 |
| Student 2 | 1 | 2 | 3 | 4 |
| Student 3 | 4 | 3 | 1 | 2 |
| Student 4 | 1 | 2 | 3 | 4 |
| Student 5 | 4 | 3 | 2 | 1 |
| Student 6 | 3 | 4 | 2 | 1 |
| Student 7 | 1 | 3 | 2 | 4 |
| Student 8 | 1 | 2 | 4 | 3 |
| Student 9 | 1 | 2 | 4 | 3 |
| Student 10 | 3 | 2 | 4 | 1 |
| Student 11 | 4 | 2 | 3 | 1 |
| Student 12 | 1 | 2 | 3 | 4 |
| Student 13 | 2 | 4 | 3 | 1 |
| Student 14 | 1 | 2 | 3 | 4 |
| Manager 1 | 3 | 2 | 1 | 4 |
| Manager 2 | 2 | 3 | 4 | 1 |
| Manager 3 | 1 | 3 | 4 | 2 |
| Manager 4 | 2 | 4 | 1 | 3 |
| Manager 5 | 4 | 2 | 3 | 1 |
| Manager 6 | 2 | 4 | 3 | 1 |

TABLE 7
Stakeholder ranking of plans for the four plans of Table 5 calculated for Capacity = 367.4. All plans were evaluated from dissatisfaction perspective.

| Stakeholder | Plan 1 | Plan 2 | Plan 3 | Plan 4 |
|---|---|---|---|---|
| Student 1 | 4 | 1 | 2 | 3 |
| Student 2 | 1 | 2 | 3 | 4 |
| Student 3 | 2 | 3 | 1 | 4 |
| Student 4 | 1 | 2 | 3 | 4 |
| Student 5 | 4 | 3 | 2 | 1 |
| Student 6 | 3 | 4 | 2 | 1 |
| Student 7 | 1 | 2 | 3 | 4 |
| Student 8 | 1 | 2 | 4 | 3 |
| Student 9 | 1 | 1 | 1 | 1 |
| Student 10 | 1 | 3 | 1 | 4 |
| Student 11 | 1 | 2 | 4 | 3 |
| Student 12 | 1 | 2 | 3 | 4 |
| Student 13 | 3 | 1 | 2 | 4 |
| Student 14 | 1 | 2 | 3 | 4 |
| Manager 1 | 2 | 3 | 1 | 4 |
| Manager 2 | 3 | 2 | 4 | 1 |
| Manager 3 | 4 | 3 | 2 | 1 |
| Manager 4 | 1 | 2 | 3 | 4 |
| Manager 5 | 4 | 3 | 2 | 1 |
| Manager 6 | 3 | 1 | 2 | 4 |

included 20 stakeholders. Six of the stakeholders were from the case study company, providing additional arguments for justification of their choices. The other 14 participants were from the pool of graduate students having conducted the original Kano feature evaluation before. We invited all the 24 graduate students who participated in the feature evaluation. 22 of them also participated in the second round of evaluation. However, only 14 graduate students answered the questions completely. Our survey was two staged:

**Phase 1:** We asked all the 20 participants to rank the four SDO plans generated for the capacity level of $cap(1) = 367.4$. They were asked to do this from both satisfaction and dissatisfaction perspective.

**Phase 2:** We asked the six managers to plan based on the 36 features. They manually planned based on perceived value of features considering $cap(1) = 367.4$.

#### 5.5.1 Phase 1: Ranking of plans

In Phase 1 of the survey we asked stakeholders to rank the four SDO generated plans. The results of ranking are reported in Table 6 (satisfaction perspective) and Table 7 (dissatisfaction). Using Fleiss Kappa test [67] for measuring inter-rater agreement showed a slight to poor agreement between the 20 participants. Considering the ranks assigned from satisfaction perspective, the Kappa value is 0.0409 ($p\ value$ = 0.05). Considering dissatisfaction, the Kappa value is 0.0649 ($p\ value$ = 0.002). Both values indicate a low agreement. This means, that there is no obvious preference among the plans, or stated alternatively, each of the four plans is considered important by at least some of the stakeholders.

By comparing plans per stakeholder, among stakeholders and between criteria, we found that:

- **One plan does not fit all**: For both planning objectives, there is substantial variation between stakeholders in terms of what they consider their preferred solution,
- **One symmetric criterion is not enough**: Six of the 20 stakeholders have a varying top preference when comparing plans selected from satisfaction and dissatisfaction perspective.

#### 5.5.2 Phase 2: Comparison with manual plans

We asked six project managers to propose their manual plan. They were offered all the 36 candidate features, their effort estimates and the total effort available. The resulting plans are shown in Table 8.

From analyzing these manual plans, we observed that the satisfaction and dissatisfaction of each plan are far below the quality of the SDO generated plans (they generate lower satisfaction and higher dissatisfaction). We compared the six manual solutions with the four optimized plans. On average, SDO solutions perform 59.2% better in terms of satisfaction and 83.4% in terms of (avoided) dissatisfaction. Looking into the case of capacity of 367.4, there are 14 core features suggested by all optimized plans. These features are *F1, F2, F3, F9, F11, F14, F15, F17, F18, F19, F21, F22, F25, F30*. The manual plans on average offered only 8 of the core features (ranking between 5 and 10). In other words, on average, each of the manual plans are missing six of the core features that have found essential for maximizing satisfaction and minimizing dissatisfaction. This is considered critical as the absence of core features is expected to have a negative impact on the success of the product release.

### 5.6 Threats to validity

The main goal of our case study is in evaluating the hypotheses that (i) trade-off release plans generated by SDO

TABLE 8
Manual plans generated by company stakeholders M1 to M6, calculated for Capacity = 367.4.

| Manager | Feature set | Satisfaction | Dissatisfaction |
|---|---|---|---|
| **M1** | F2,F4,F5,F8,F9,F10,F13,F15,F17,F19,F29 | 6.50 | 8.39 |
| **M2** | F2,F3,F4,F9,F10,F11,F15,F19,F21,F30,F31,F34,F35 | 6.16 | 8.25 |
| **M3** | F1,F3,F4,F5,F9,F10,F11,F12,F16,F17,F18,F19,F21,F22,F24,F27 | 7.22 | 7.78 |
| **M4** | F1,F3,F4,F5,F9,F10,F11,F15,F19,F21,F28,F31,F34,F35 | 6.61 | 8.03 |
| **M5** | F3,F4,F5,F6,F8,F9,F10,F13,F14,F15,F16,F20,F28 | 6.53 | 8.01 |
| **M6** | F1,F2,F4,F5,F8,F9,F10,F12,F13,F15,F20,F28 | 5.80 | 8.72 |

perform better than individual plans generated by random and heuristic search and (ii) are considered valuable by stakeholders.

How trustworthy are the results gained and what threats still exist? Following [63], we discuss four types of threats. To further increase credibility, we have provided access to all case study data. Related to reliability of results, most of them are based on objective measures which have been used as well in other studies in the context of software engineering multi-criteria decision-making (compare e.g., [73], [80]).

### 5.6.1 Construct validity

How valuable is comparison with random and heuristic search? Random search has been used for various studies in software engineering for the same purpose since it was suggested by Acuri and Briand [3]. Greedy heuristics, in general, are widely used as a light-weight technique in decision-making, and the same is true in release planning [62].

Another construct validity aspect is if stakeholders got the right understanding of the survey questions and in generating manual plans. To mitigate this risks, we gave a 10 minutes instruction and description for filling a survey over the phone. All the candidate features and their effort were made available to them. As they were familiar with the content of the features, their decisions are considered to be well justified.

### 5.6.2 Internal validity

There are no causal relationship statements made on the planning results studied in the case study. The different plans compared were generated by the various techniques without further impacting (confounding) factor. We used a real-world project for case study evaluation. Also, the survey results were not impacted by any other obvious factor.

### 5.6.3 External validity

Following the classification of [78] for generalizing software engineering theories, our case study is what is called a *Lab-to-field generalization*. As a form of technology validation, it provides arguments for the suitability and usefulness of the asymmetric planning method. Survey Phase 1 evaluated whether having a set of trade-off solutions is valuable. Comparison between the 20 stakeholders indicates that there are subjective differences among stakeholder perspectives and preferences. Ignoring dissatisfaction would imply the risk of stakeholder dissatisfaction. Even though the number of participants in Phase 2 of the survey is small, most manager stakeholders believe the method is useful and scalable. Five out of 6 stakeholders strongly agree on the value of optimality as achieved by the SDO method. Because of all that, we expect that the case study gives positive signals for the external validity of the method.

### 5.6.4 Reliability validity

Students served as 14 (out of 20 in total) stakeholders evaluating the value and attractiveness of features. Most of them were familiar with OTT services such as Netflix. In-line with the results of [8], we consider them as suitable stakeholders for case study analysis. Furthermore, stakeholders' importance was decided based on their (self-evaluated) familiarity with the context. In other words, if a student did not feel familiar with OTT services, her impact on the actual results was low. Previous studies showed that the use of students for evaluation in the industrial context is fine when using with caution about limitations and threats [69].

## 6 DISCUSSION

From performing the case study, we can make some conclusions about the proposed SDO approach. We approached the six project managers to perform a more in-depth analysis of SDO plans. We asked these stakeholders to answer eight questions related to different aspects of the usefulness of plans and different aspects of SDO. The questions were defined on a five-point scale with 1 to 5 being *very low* to *very strong* agreement. The categories of questions and the manager's responses are summarized in Table 9.

### 6.1 Relevance of ARP

Providing the best set of features for upcoming releases is decisive for product success. There is a risk of providing features that are not relevant as well as missing requested features [62]. The asymmetry between satisfaction and dissatisfaction was confirmed by literature [42]. Looking into the survey results (Tables 7 and 8), it became clear that different plans are preferred between stakeholders. Also, when comparing the two plans preferred per stakeholder (one for satisfaction, one for dissatisfaction), again, there are differences in about one-third of the cases.

TABLE 9
Manager's level of agreement to Phase 2 survey questions

| Question | M1 | M2 | M3 | M4 | M5 | M6 |
|---|---|---|---|---|---|---|
| **Usefulness of Kano** | 5 | 5 | 5 | 5 | 4 | 4 |
| **Mitigating risk of offering wrong product** | 5 | 4 | 5 | 5 | 5 | 4 |
| **Efficiency of Kano** | 4 | 3 | 5 | 4 | 4 | 5 |
| **Relevance of ARP** | 5 | 4 | 5 | 5 | 5 | 4 |
| **Value of diversity** | 5 | 5 | 5 | 5 | 5 | 5 |
| **Optimized solution vs. ad hoc** | 5 | 5 | 5 | 5 | 4 | 5 |
| **Optimized solution vs. heuristic** | 4 | 5 | 4 | 5 | 5 | 5 |
| **Scalability of SDO** | 3 | 5 | 4 | 3 | 5 | 5 |

We asked the six managers in our case study about the asymmetry between satisfaction and dissatisfaction.

> To what extent do you agree that satisfaction and dissatisfaction both are relevant for planning product releases and need to be considered in conjunction?
> *All participants either agreed or strongly agreed that both criteria are relevant and need to be considered in conjunction, which is the key idea of ARP (See Tables 7 and 8).*

Analyzing the different features from both satisfaction and dissatisfaction perspectives and including all stakeholders in this process helps to not miss essential features.

> To what extend do you agree that a systematic planning method which includes stakeholder opinions related to both satisfaction and dissatisfaction reduces the risks of developing the "wrong" product?
> *Four developers were strongly agreed, and two agreed that joint consideration of satisfaction and dissatisfaction helps mitigate the risk of developing the wrong product.*

## 6.2 Usefulness of Kano for measuring feature satisfaction and dissatisfaction

We introduced three methods for measuring satisfaction and dissatisfaction, and the ARP model works independently of which of them is used. We applied the continuous Kano model for our case study as it has been widely discussed in literature [74]. Offering the results of the case study, we asked the managers about the usefulness of Kano model and its efficiency.

> To what extent do you think that Kano evaluation was used to understand users better?
> *Four managers strongly agreed, and two managers agreed that the Kano was useful for their company.*
>
> To what extend do you think the additional effort (from answering ten questions per feature based on Kano) is worthwhile?
> *Two managers strongly agreed, three agreed, and one manager was neutral about the efficiency of the Kano model.*

While the literature confirms and managers agree with the value of the Kano model, the usefulness and efficiency of it are rather context specific and alternative methods should be evaluated.

## 6.3 Value of diversity

The diversity principle formulated in [62] says that *A single solution of a cognitive complex problem is less likely to reflect the real-world problem solving needs when compared to a portfolio of qualified solutions that are structurally diversified*. In its nature, solutions generated from SDO are trade-off balancing satisfaction against dissatisfaction. The plans differ in their structure.

We illustrate this argument by computing the symmetric differences between all pairs of SDO generated plans for the case of $Cap(1) = 367.4$. The symmetric difference between two sets is the set of elements that are in either of the sets, but not in the intersection. We can see that the minimum number of elements in the symmetric difference is two, and the maximum number is five. The results are presented as Table 10. With a set of 14 features being offered for all plans, there is variation in the remaining features.

ARP includes the provision of alternative solutions, balancing between satisfaction and dissatisfaction. As can be seen from the inter-rater agreement in terms of ranking plans, there is a slight to poor agreement between the 20 participants of Phase 1 of the survey. The survey results show different preference of plans between stakeholders and between the two planning objectives.

This confirms the value of offering a set of alternative solutions instead of prescribing just one plan. The deeper reason for this is that any problem description can never be complete, and the personal preference naturally varies between stakeholders because of the different degree of tacit knowledge they have about the problem.

> How valuable do you consider having several alternative plans as the result of SDO (Pareto solutions) as opposed to offering just one?
> *All the six managers strongly agreed that diversity of solutions is desirable.*

## 6.4 Value of optimality

We compared the results of SDO with (i) random search, (ii) heuristic search, and (iii) manual plans generated by stakeholders of the case study company. All these comparisons showed that the solutions generated by SDO are better. The value of optimized plans in comparison with the heuristic of ad hoc plans was confirmed by all stakeholders, as they almost all strongly agreed on that.

> How valuable do you consider optimized solutions in comparison to solutions generated ad hoc, based on gut feeling?
> *Five managers were strongly agreed, and one agreed that optimized solutions are better than ad hoc solutions based on gut feelings.*
>
> How valuable do you consider optimized solutions in comparison to solutions generated heuristically?
> *Four managers strongly agreed, and two agreed that optimized solutions are better than solutions generated heuristically.*

TABLE 10
Symmetric differences between the four SDO generated plans of Table 5 calculated for Capacity = 367.4.

| Plan | Plan 2 | Plan 3 | Plan 4 |
|---|---|---|---|
| Plan 1 | F23, F24 | F10, F23, F24, F36 | F4, F6, F10, F24, F36 |
| Plan 2 | * | F10, F36 | F4, F6, F10, F23, F36 |
| Plan 3 | * | * | F4, F6, F23, F36 |

### 6.5 Scalability

Generation of optimized trade-off plans for the advanced model of ARP does not come for free: The method requires feature evaluation and information gathering. In total, the ARP formulation increases the conceptual complexity of the task of the decision-maker. Depending on their needs, they can select from different modeling options for getting the information needed. We have provided alternative techniques to elicit the relative value of features related to satisfaction respectively dissatisfaction of a release.

Even though the computational complexity of the problem is NP-hard, Gurobi is able to solve benchmark integer linear programming problems[5] with several thousands of variables with proven optimality.

The computation time for SDO is comparable with the computation time of the random search and the heuristic solutions. Using an Intel(R) core i7-2620M CPU@ 2.70 GHZ computer, using a random search (Algorithm 1) for the problem in our case study took 48ms, while heuristics (Algorithm 2) took 55ms on average to find a solution. SDO finds a solution in 54ms per iteration.

> To what extend the size and complexity of the product impact the performance of the SDO approach?
> *Three managers strongly agreed, one agreed and two were neutral about the scalability of SDO.*

### 6.6 Limitations

The formulation of release planning has aspects of wickedness [50]. That means, as for other design problems, there is no ultimate answer on what is a right model. Robertson and Robertson [61] showed that stakeholders do value gains and losses unequally. A few models were provided relying on the asymmetry between satisfaction and dissatisfaction such as Kano analysis and prospect theory [71].

When SDO should not be applied? The whole investment into advanced planning methods like the proposed method SDO does not make sense when (i) the planning problem is not complex in terms of number of features, number of stakeholders involved, and number of competitors on the market, (ii) strategic perspective of planning (as opposed to short term considerations and/or continuous delivery) is not valuable as there is too much change in market conditions and data used for performing the planning process, (iii) the organization is not mature enough to provide qualified input for the planning process, and (iv) the differentiation between feature and product satisfaction and dissatisfaction is not important.

### 7 RELATED WORK

In this section we discuss the asymmetry between customers' satisfaction and dissatisfaction referring to other disciplines and evaluate how they were used in software engineering methods. We also give a brief overview on symmetric release planning methods and the use of multi-objective optimization in software engineering [82].

5. http://plato.asu.edu/ftp/milpc_log1/benchmark.gurobi.out

### 7.1 Satisfaction and Dissatisfaction in Prioritization and Planning

Several models are proposed for understanding customers' satisfaction and dissatisfaction. Satisfaction factors are classified differently mainly followed by the Kano model. These models in general model customers reaction to the product changes in a nonlinear and asymmetric model. Kano et al. [30] suggested a questionnaire which includes two questions related to each feature. The functional question evaluates the customers reaction to the presence of a particular feature. The dysfunctional question assesses the customers reaction in the absence of the same feature. Cadotte & Turgeon [14] suggested a model with four different feature considering the potential of features for raising positive or negative feelings. They defined qualitative levels for identifying typical features for each category. Anderson and Mittal [57], Backhause and Bauer [57], Bitner et al. [57], and Brandt [57] and several others proposed methods for modeling the asymmetry between customers satisfaction and dissatisfaction.

The very first Kano model was introduced in the early 1980s [30]. Mikulic [43] classifies the Kano technique as highly reliable for classification of quality attributes at the design stage. The use of this model for requirement prioritization in non-software products was elaborated [30], [66]. The results showed that the requirements have an asymmetric effect on stakeholder satisfaction and dissatisfaction.

The lack of quantitative assessment in the traditional Kano model limits the value of the decision support provided [79]. As a result of that, several Kano extensions have been proposed. Violante et al. [74] found ten extensions of the original Kano method. Berger's quantitative model [11] (As we used in this paper) is the most simplified and earliest proposed quantitative model.

Another critique of the traditional Kano model is its limitation in precisely defining fulfillment and non-fulfillment of stakeholder satisfaction. The traditional Kano model categorizes each feature into exactly one Kano category. This does not sufficiently reflect the complex sentiments of an individual [34], [43]. Lee et al. [34] proposed the fuzzy Kano questionnaire which extends the traditional approach by providing a questionnaire to help stakeholders explain their feelings. The fuzzy Kano model uses a membership function and assigns numeric degrees to stakeholders feelings. While fuzzy Kano and continuous Kano create the same type of information, the difference in the continuous Kano is that information is also used for subsequent release planning purposes.

Software products and features have numerous value constructs [32]. Khurum et al. [32] provided a comprehensive value map for software. Acquiring values for comprehensive feature prioritization needs careful analysis [32]. For these reasons, feature prioritization methods and, even further, release planning approaches have assumed that priorities are coming from stakeholders.

A spectrum of the prioritization techniques is discussed in literature [1], [10], [22], [59]. Prioritization techniques have been widely discussed in software engineering within release planning methods or independently. An overview of these approaches was given by Berander and Andrews

TABLE 11
Consideration of satisfaction and dissatisfaction as prioritization criteria among the 137 studies systematically reviewed by [1] and [32]. Papers not referenced in the table neither consider satisfaction nor dissatisfaction.

| Study | Satisfaction as prioritization criteria | Dissatisfaction as prioritization criteria | Description |
|---|---|---|---|
| Voola and Babu [76] | Considered | - | - |
| Liu et al. [37] | Considered | - | - |
| Raharjo et al. [55] | Considered | - | - |
| Kukreja [33] | Considered | - | - |
| Voola and Babu [75] | Considered | - | - |
| Forouzani et al. [23] | Considered | - | - |
| Babar et al. [5] | Considered | - | - |
| Otero et al. [51] | Considered | - | - |
| Gaur et al. [25] | Considered | - | - |
| Carod and Cechich [16] [15] | Considered | - | - |
| Daneva and Herrmann [19] | - | Considered | - |
| Liu et al. [36] | Considered | - | - |
| Karlsson [31] | Considered | - | - |
| Fehlmann [21] | Considered | Considered | Uses Kano but does not refer to dissatisfaction. Instead, considers satisfaction and technical excellence. |
| Pitangueira et al. [52] | Considered | - | - |
| Lehtola and Kauppinen [35] | Considered | Considered | Points to Kano model as a basic model for prioritization. |
| Berander and Andrews [10] | Considered | - | - |
| Samer et al. [44] | Considered | - | - |
| Botta and Bahill [12] | Considered | - | - |
| Hu et al. [28] | Considered | - | - |
| Barney et al. [6] | Considered | - | - |
| Berander [9] | Considered | - | - |
| Robertson and Robertson [61] | Considered | Considered | Their results showed that stakeholder satisfaction of receiving and dissatisfaction of not receiving a feature is not equal. |
| Ziemer et al. [81] | Considered | - | - |
| Akker et al. [72] | Considered | - | - |
| Regnell et al. [58] | Considered | - | - |
| Benestad et al. [7] | Considered | - | - |
| Racheva [54] | Considered | - | - |
| Barney et al. [6] | Considered | - | - |
| Aurum and Wohlin [4] | Considered | - | - |
| Logue and Kevin [38] | Considered | - | - |
| Ruhe et al. [26], [62], [65] | Considered | Considered | Introduced EVOLVE family optimizing a linear combination (single criterion) of planning criteria. |

[10]. Achimugu et al. [1] discovered 49 distinct prioritization techniques and reported AHP [64] as the technique with the highest citation and utilization. Feature prioritization is possible based on different criteria. On the other side, the study by Riegel and Doerr [59] classified stakeholder satisfaction and stakeholder dissatisfaction among the top ten prioritization criteria [59].

The missing phenomenon of interest for our study within currently available literature reviews [1], [10], [22], [59] is the extent of conjoint consideration of satisfaction and dissatisfaction for feature prioritization and release planning. To gather this information, we took the 73 papers selected by Achimugu et al. [1] for their systematic literature study and the 83 papers analyzed by Riegel and Doerr [59]. These two studies had 15 papers in common. As a result, we analyzed 141 papers to find if and how satisfaction and dissatisfaction were considered jointly in feature prioritization and release planning methods. The results of this analysis are reported in Table 11.

We found that 33 studies considered satisfaction or dissatisfaction as their planning criteria. However, only four of these studies jointly considered satisfaction and dissatisfaction. Among these four studies, Fehlmann [21] and Lehtola and Kauppinen [35] refer to the Kano model for feature prioritization, but without entering into release decision making. Robertson and Robertson [61] showed that stakeholder values gains and losses are unequal and stakeholder satisfaction of receiving a feature is not equal to the dissatisfaction of not receiving that feature. For release planning, EVOLVE was introduced by Ruhe et al. [26], [65] as a method being flexible in the number and selection of planning criteria.

In software release planning, it has mostly been assumed that receiving a feature provides specific stakeholder satisfaction which is equal to stakeholder dissatisfaction from not receiving that feature. In other words, release planning methods assumed that satisfaction and dissatisfaction are symmetric. However, the Kano model [30] demonstrated that stakeholders satisfaction and dissatisfaction occurs conjointly. Our proposed ARP formulation points to the essence of conjoint consideration of satisfaction and dissatisfaction in release planning because using one of these values without the other one will result in biased and incomplete results.

## 7.2 Multi-objective optimization in software engineering

Consideration of multiple objectives is a growing trend in software engineering. One criterion (such as cost, revenue, time-to-market) only provides a partial direction for determining best strategies. The notion of Pareto-optimality guides searches towards solutions that can only be improved towards one criterion by compromising against another one. Aligned with the trend towards more analytics and subsequent quantitative investigations, multi-objective optimization has been recently applied to the Next Release Problem [73] [53], to software architectures [56], re-factoring [41], requirements selection [20], and model merging [40]. A more general release problem with looking at three releases ahead and having revenue and cost as optimization criteria and objectives were studied in [80]. With the exception of [73], all these approaches were based on evolutionary and search-based techniques. In [80], an empirical study of meta- and hyper-heuristic search was performed for a series of ten real-world data sets. The authors found that hyper heuristics [13] were most successful.

For multi-objective release planning, it was shown in [73] that ILP is applicable and competitive with search-based algorithms in terms of computational effort. A similar result was obtained already earlier for the case of single criterion release planning by van den Akker et al. [72]. ILP results are even better with regards to the guaranteed optimality of the solutions obtained. Based on that finding, we have continued that route and have applied ILP on a more general and broader class of bi-objective release planning problems which are based on asymmetric performance of features.

Integer linear programming was used by Veerapen et al. [73] to solve the single and bi-objective Next Release Problem. While there has been a dominance of search-based techniques in the past (starting with the genetic algorithm of Greer and Ruhe [26]), the authors have shown that integer linear programming-based out-performs the NSGA-II genetic approach on large bi-objective instances.

## 8 Summary and Conclusions

Release planning is a cognitively and computationally complex problem. From improving the information used in the planning process, we have a better chance to address the right problem. We proposed a bi-criteria problem formulation to determine a set of trade-off release planning solutions. From applying the linear integer programming based method SDO, a set of optimized trade-off solutions can be determined. The practical usefulness of the approach was demonstrated by a case study taken from the context of an app store market. We consider the method ready to be transitioned into practice: Besides the general importance of a systematic and objective method for deciding future product releases, SDO for the first time looks into the asymmetry between satisfaction and dissatisfaction. This increases the validity of the underlying model for capturing customer perspective on feature value. From a return-on-investment (ROI) perspective: More comprehensive data and knowledge is needed to run SDO and leverage its results. This is compensated by helping software product owners to reduce the risk of unnecessary features or satisfying some customers while making others extremely dissatisfied. Further work is needed to evaluate the hypothesis that the benefit of SDO is the higher, the more mature the processes are in the organization, the more users and customers the product has, and the more competitive the space of the product is.

We foresee a wider usage and future research on looking at both satisfaction and dissatisfaction beyond its consideration in release planning. The same idea is applicable for all types of prioritization and planning, and for all types of software life-cycle models. Following the arguments of Stol and Fitzgerald [68], we see our paper as a contribution to software engineering theory. Among their formulated domains of their (Research Path Schema), we classify our work as a contribution to the conceptual domain. The conceptual contribution is based on the key principle that (i) satisfaction and dissatisfaction need to be treated in conjunction, (ii) their relationship is asymmetric, and (iii) treating the relationship as a trade-off between the two criteria.

Our proposed ARP model can be based on the (continuous) Kano model for feature needs elicitation. Another model which could alternatively be used for asymmetric planning is the prospect theory as proposed by Tversky and Kahneman [71]. Prospect theory is emphasizing the difference between valuing gain and loss and how stakeholders would value decisions that are made based on gain rather than on loss.

After the initial case study, the method is ready for more comprehensive industrial evaluation. A more comprehensive empirical evaluation is required to validate the usefulness of the method over the "traditional" (symmetric) approaches. In particular, the additional effort needed to solicit more comprehensive stakeholder information and to perform the satisfaction-dissatisfaction optimization needs to be related to the added value gained from having a set of optimized release plan alternatives.

## Acknowledgments

We thank all the anonymous reviewers and the Associate editor for their valuable comments and suggestions. We wish to thank Mohsen Ansari, Navid Pourmomtaz, and Maryam Soleimani for their help on the implementation of this research. We are grateful to Des Greer for helpful discussions on a former version of the paper. This research was partially supported by the Natural Sciences and Engineering Research Council of Canada, NSERC Discovery Grant 250343-12.

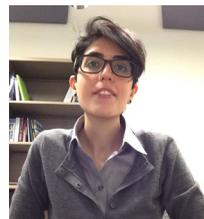

**Maleknaz Nayebi** is a Postdoctoral fellow at the University of Toronto. She got her PhD at the Software Engineering Decision Support lab from The University of Calgary in Canada. The PhD was on *Analytical Release Management for Mobile Apps*. She has six years of professional software engineering experience. Her main research interests are in mining software repositories, release engineering, open innovation and empirical software engineering. Maleknaz co-chaired RE data track 2018, IWSPM 2018, IAS-ESE 2018 advanced school, and OISE 2015. Maleknaz is a student member of the IEEE and ACM.

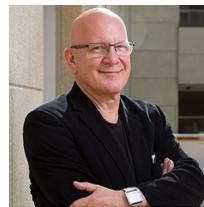

**Guenther Ruhe** holds an Industrial Research Chair in Software Engineering at University of Calgary. Dr. Ruhe received a doctorate habil. nat. degree (Computer Science) from University of Kaiserslautern. From 1996 until 2001, he was the deputy director of the Fraunhofer Institute for Experimental Software Engineering FhIESE. Since 2016, he serves as the Editor in Chief of the journal of Information and Software Technol- ogy, published by Elsevier. His main research interests are in the areas of Product Release Planning, Software Project Management, Decision Support, Data Analytics, Empirical Software Engineering as well as Search-based Software Engineering. He is a Senior member of IEEE and a member of the ACM. Dr. Ruhe is the Founder and CEO of Expert Decisions Inc., a University of Calgary spin-off company created in 2003.

# APPENDIX I: Sample Feature Attribute Calculation from Applying Continuous Kano

As an example for feature value prediction, continuous Kano attribute calculation for feature F15 is performed. From both Figure 3 and Figure 4, we observe that there is a clear differentiation between Kano attributes per feature. The same is true for the satisfaction-dissatisfaction profile between features. This is the purpose of the fuzzy Kano model outlined in Section 3: To extract the complex and multi-dimensional customer value perception of features. For example, F25 is primarily seen as a must have (M) feature, but also has degrees of O, A, and I attributes as well.

For the sample feature F15, we describe the steps of continuous Kano analysis using the actual stakeholder input received.

**Step 1: Kano questionnaire**

*Electronic Program Guide (EPG) is a continuously updating menu displaying the broadcast programming or scheduling information for current and upcoming programs.*

Functional question: **How would you feel if "Displaying Electronic program guide (EPG)" was provided with this mobile app?**

___% I like it that way
___% It must be that way
___% I'm indifferent
___% I can live with it that way
___% I dislike it that way

Dysfunctional question: **How would you feel if "EPG" was not provided with this mobile app?**

___% I like it that way
___% It must be that way
___% I'm indifferent
___% I can live with it that way
___% I dislike it that way

**Step 2: Calculating Kano attribute scores for each feature**

In Step 2 of the process, we used Equations (3) to (6) to calculate the Kano attributes of feature F15. For example, Stakeholder 5 stated "Like" 100% for the inclusion of F15 (functional question) and 5% "Neutral", 11% "Live with", and 84% "Dislike" about not including F15 (dysfunctional question). In that case, Table 3 (for this stakeholder and for F15) looks as below:

| F15 evaluated by Stakeholder 5 | | Dysfunctional questions | | | | |
|---|---|---|---|---|---|---|
| | | Like | Must-be | Neutral | Live with | Dislike |
| **Functional questions** | Like | 0 | 0 | 0.05 | 0.11 | 0.84 |
| | Must-be | 0 | 0 | 0 | 0 | 0 |
| | Neutral | 0 | 0 | 0 | 0 | 0 |
| | Live with | 0 | 0 | 0 | 0 | 0 |
| | Dislike | 0 | 0 | 0 | 0 | 0 |

We apply Equations (3) and (4) with the data summarized in the above table, resulting in:

score$_A$(25, 5) = ($U_i \times D_{ii}$) + ($U_i \times D_{iii}$) + ($U_i \times D_{iv}$) = (1 × 0) + (1 × 0.05) + (1 × 0.11) = 0.16

score$_O$(25, 5) = ($U_i \times D_v$) = (1 × 0.84) = 0.84

Note that we skipped calculations of *Score$_M$*, and *Score$_I$* for Stakeholder 5 as scores were equal to zero. The results of both computations obtained for all stakeholders are summarized in Step 2 with the table below. Results for Stakeholder 5 as just explained are highlighted.

**STEP 2 - Calculating Kano attributes for feature F15 (sum of rows might be unequal to one because of R attribute)**

| Stakeholder | M | O | A | I |
|---|---|---|---|---|
| Stakeholder 1 | 1 | 0 | 0 | 0 |
| Stakeholder 2 | 0 | 0 | 1 | 0 |
| Stakeholder 3 | 0 | 0 | 0.51 | 0.49 |
| Stakeholder 4 | 1 | 0 | 0 | 0 |
| Stakeholder 5 | 0 | 0.84 | 0.16 | 0 |
| Stakeholder 6 | 0.18 | 0.73 | 0.07 | 0.01 |
| Stakeholder 7 | 0 | 0 | 0 | 1 |
| Stakeholder 8 | 0 | 0 | 0.49 | 0.51 |
| Stakeholder 9 | 0.470 | 0.090 | 0.04 | 0.36 |
| Stakeholder 10 | 0 | 0 | 0 | 1 |
| Stakeholder 11 | 1 | 0 | 0 | 0 |
| Stakeholder 12 | 1 | 0 | 0 | 0 |
| Stakeholder 13 | 0 | 0 | 1 | 0 |
| Stakeholder 14 | 0 | 0 | 0 | 1 |
| Stakeholder 15 | 0 | 0 | 0 | 1 |
| Stakeholder 16 | 0.187 | 0.053 | 0.167 | 0.59 |
| Stakeholder 17 | 1 | 0 | 0 | 0 |
| Stakeholder 18 | 0.41 | 0 | 0 | 0.59 |
| Stakeholder 19 | 0.006 | 0.0744 | 0.856 | 0.06 |
| Stakeholder 20 | 1 | 0 | 0 | 0 |
| Stakeholder 21 | 0.272 | 0.408 | 0.192 | 0.12 |
| Stakeholder 22 | 0.66 | 0.34 | 0 | 0 |
| Stakeholder 23 | 0 | 1 | 0 | 0 |
| Stakeholder 24 | 0.25 | 0.25 | 0.25 | 0.25 |

**STEP 3 - Calculating weighted averages by aggregating stakeholders score**

| Stakeholder | Weight | M | O | A | I |
|---|---|---|---|---|---|
| Stakeholder 1 | 1 | 1 | 0 | 0 | 0 |
| Stakeholder 2 | 6 | 0 | 0 | 6 | 0 |
| Stakeholder 3 | 8 | 0 | 0 | 4.08 | 3.92 |
| Stakeholder 4 | 8 | 8 | 0 | 0 | 0 |
| Stakeholder 5 | 6 | 0 | 5.04 | 0.96 | 0 |
| Stakeholder 6 | 9 | 1.62 | 6.57 | 0.63 | 0.09 |
| Stakeholder 7 | 8 | 0 | 0 | 0 | 8 |
| Stakeholder 8 | 8 | 0 | 0 | 3.92 | 4.08 |
| Stakeholder 9 | 9 | 4.23 | 0.81 | 0.36 | 3.24 |
| Stakeholder 10 | 1 | 0 | 0 | 0 | 1 |
| Stakeholder 11 | 2 | 2 | 0 | 0 | 0 |
| Stakeholder 12 | 8 | 8 | 0 | 0 | 0 |
| Stakeholder 13 | 8 | 0 | 0 | 8 | 0 |
| Stakeholder 14 | 1 | 0 | 0 | 0 | 1 |
| Stakeholder 15 | 3 | 0 | 0 | 0 | 3 |
| Stakeholder 16 | 3 | 0.561 | 0.159 | 0.501 | 1.77 |
| Stakeholder 17 | 8 | 8 | 0 | 0 | 0 |
| Stakeholder 18 | 8 | 3.28 | 0 | 0 | 4.72 |
| Stakeholder 19 | 8 | 0.048 | 0.595 | 6.848 | 0.48 |
| Stakeholder 20 | 3 | 3 | 0 | 0 | 0 |
| Stakeholder 21 | 9 | 2.43 | 3.69 | 1.728 | 1.08 |
| Stakeholder 22 | 3 | 1.98 | 1.02 | 0 | 0 |
| Stakeholder 23 | 1 | 0 | 1 | 0 | 0 |
| Stakeholder 24 | 3 | 0.75 | 0.75 | 0.75 | 0.75 |
| **Weighted average** | | 0.332 | 0.148 | 0.258 | 0.253 |

Having stakeholders' Kano evaluation for F15 and the weight of each stakeholder, in Step 3' we calculate the weighted average of Kano attributes for F15.

**Step 4 Calculating feature satisfaction and dissatisfaction values**

Following Equations (8) and (9), we calculate the satisfaction that is expected among stakeholders from obtaining F15 and the dissatisfaction that may be caused if F15 were not offered:

$$S(15) = \frac{F_A(15) + F_O(15)}{F_A(15) + F_O(15) + F_M(15) + F_I(15)} = \frac{0.258 + 0.148}{0.258 + 0.148 + 0.332 + 0.253} = 0.406$$

$$DS(15) = \frac{F_M(15) + F_O(15)}{F_A(15) + F_O(15) + F_M(15) + F_I(15)} = \frac{0.332 + 0.148}{0.258 + 0.148 + 0.332 + 0.253} = 0.48$$

# APPENDIX II: SDO Tool Set-up and Application

We developed our SDO solutions using `MATLAB` interface for the `Gurobi` optimizer. In this appendix we show how to setup the environment and how to use SDO to plan for two releases (K=2) with the data of our case study. The code snippets for SDO is presented as the complementary material along with the paper. To use the code, the `Gurobi` environment should be set up in `MATLAB`. This appendix provide a brief guideline on how to set up the environment for SDO.

Step 1: Install latest version of `MATLAB`.

Step 2: Install `Gurobi`.

Step 3: Setup `Gurobi` within `MATLAB`
Open the `MATLAB` software and on the left window with folder explorer go to the directory where `Gurobi` is installed like ``C:\gurobi605\win64\matlab"``. There are several files inside the folder as shown below:

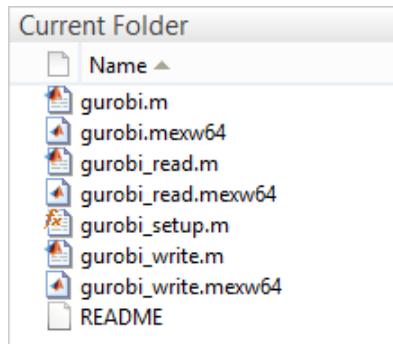

Then, on the command window write: `gurobi_setup`. If the setup would be successful the below would appear in your command window.

Step 4: Running SDO snippets. Place the SDO scripts in your preferable directories. With the folder explorer on left window of MATLAB go to the directory where project files are located. Run `SDO.m`.

The provided snippets have two parts: Part 1, that solves one bi-objective problem considering satisfaction and dissatisfaction. Also, Part 2, that varies alpha values and calls the optimizer several times. To run the program, you only need to execute SDO.m which is the implementation of Part 2 iteratively calling Part 1. Below algorithm shows the pseudo code of this implementation.

**Algorithm 1** SDO implementation
```
for i = 1:1000 do
    for for j=1:numberoffeatures do
        model.obj(j) = alpha*satisfaction(j) + (1-alpha)*dissatisfaction(j);
    result = gurobi(model, params);
    totaleffort = 0;
    totalsatisfaction = 0;
    totaldissatisfaction = all feature dissatisfaction;
    SingleObjNextRelease;    #Part1
    alpha = alpha + 0.001;
} #end for
```

The SDO tool is designed based on the ARP formulation we provided in Section 3. ARP and the SDO tool can accommodate the release planning of a product with multiple stakeholders, and multiple releases. The below screen shot of the tool shows the application of the SDO tool for our case study in Section 5 (RQ3).

```
Editor - C:\Users\Maleknaz\Dropbox\PhD\TSE revision\TES_Shared with Guenther\MohsenMaryamProjectV3\final\stakeholders.m
 stakeholders.m  +
34 -     efforts = (sparse(numberofservices));
35 -  □ for i=1:numberofstakeholders
36 -         disp('##############');
37 -         effort = input ('Please enter the effort for F1:');
38 -         stakeholderweights(1,i) = input(['Please enter the weight of stakeholder ' ,num2str(i), '(1-9): ']);
39 -         totalweight = totalweight + stakeholderweights(1,i);
40 -         disp(['please enter value for each feature from stakeholder ' ,num2str(i) ,' (',...
41              num2str(numberofservices),' numbers between 0%-100%)']);
42 -  □     for j=1:numberofservices
43 -             indifferent(1, j) = input(['Indifferent value for F', num2str(j), ': ']);
44 -             reverse(1, j) = input(['Reverse value for F', num2str(j), ': ']);
45 -             musthave(1, j) = input(['Must have value for F', num2str(j), ': ']);
46 -             onedimensional(1, j) = input(['One dimensional value for F', num2str(j), ': ']);
47 -             attractive(1, j) = input(['Attractive value for F', num2str(j), ': ']);
48 -             disp('-----------');
```

```
Command Window
>> stakeholders
-------------------------------------------------------------------------------------------
| This script defines the specific of stakeholders, services and efforts                  |
| First you state the number of stakeholders, then the number of services                 |
| afterwards for stakeholder you should define it's weight(importance)                    |
| And for every stakeholder you define their value for each service                       |
| At the end you put the efforts that each service needs to be completed and maximum available effort |
-------------------------------------------------------------------------------------------
Please enter the number of stakeholders: 24
Please enter the number of features: 36
Enter number of releases:2
Enter the maximum effort capacity for Release #1:625.5
Enter discount for Release #1:0
Enter the maximum effort capacity for Release #2:140
Enter discount for Release #2:0.5
##############
Please enter the effort for F1:10.4
Please enter the weight of stakeholder 1(1-9): 8
please enter value for each feature from stakeholder 1 (36 numbers between 0%-100%)
Indifferent value for F1: 0
Reverse value for F1: 0
Must have value for F1: 95
One dimensional value for F1: 5
Attractive value for F1: 0
-----------
fx Indifferent value for F2: ...
```

As demonstrated in the above screen shot, we ran SDO for effort capacity of 367.4 as detailed in our case study in Section 5. Here, we run the same case study with 24 stakeholders and 36 features, but with the planning horizon of two releases. Release 1 has capacity 625.5, and Release 2 has a capacity of 140. In Equation (14), we use the discount factor of 0.5 for the second release.

Each stakeholder evaluated each feature following Kano model as we detailed in Section 5.1. As input, SDO gets all the Kano values (M, O, A, R, I) defined per feature and per stakeholder. Each stakeholder has a weight between one to nine which in our case study is determined from self-evaluate familiarity with release planning and OTT services.

With this input, we ran SDO to get the optimized plan over two releases. Following Equations (12) and (13), if a feature is offered in Release 1 (2) this would be indicated by "1" ("2"), respectively. If a feature is postponed (not being implemented in these two releases), this is expressed by "3".

```
Editor - C:\Users\Maleknaz\Dropbox\PhD\TSE revision\TES_Shared with Guenther\MohsenMaryamProjectV3\final\stakeholders.m
stakeholders.m
16
17      load('pureconfig');
18      atts = zeros(1000,1);
19      vals = zeros(1000,1);
20      alpha = 0.001;
21      totaleffort=0;
22
23      capacity = 30;
24      model.rhs(1) = capacity;
25      disp(['The requested capacity is:', num2str(model.rhs(1))]);
26
27      for i=1:1000
28          for j=1:numberofservices
29              model.obj(j) = alpha*satisfaction(j) + (1-alpha)*dissatisfaction(j);
30          end
31          result = gurobi(model, params);
32          totaleffort = 0;
33          totalsatisfaction = 0;
34
```

```
Command Window

Please enter the weight of stakeholder 1(1-9): 6
please enter value for each feature from stakeholder 24 (36 numbers between 0%-100%)
Indifferent value for F1: 15
Reverse value for F1: 0
Must have value for F1: 15
One dimensional value for F1: 35
Attractive value for F1: 35
---------------
--------------------------------
You see our plans soon .... we are calculating!
----------------------------------------------------------
--------------------------------------------------------------------------------
----------------------------------------------------------------------------------------------
FEATURES| F1,F2,F3,F4,F5,F6,F7,F8,F9,F10,F11,F12,F13,F14,F15,F16,F17,F18,F19,F20,F21,F22,F23,F23,F25,F26,F27,F28,F29,F30,F31,F32,F33,F34,F35,F36
PLAN #1 | 1, 1, 1, 1, 1, 1, 1, 1, 1, 1, 1, 3, 1, 1, 1, 2, 1, 1, 1, 2, 1, 1, 1, 1, 2, 3, 1, 3, 1, 2, 1, 3, 3, 1, 1
PLAN #2 | 1, 1, 1, 1, 1, 1, 1, 1, 1, 1, 1, 1, 1, 1, 3, 1, 1, 1, 1, 1, 1, 1, 3, 2, 1, 2, 1, 1, 1, 3, 3, 3, 2
PLAN #3 | 1, 1, 1, 1, 1, 1, 1, 1, 1, 1, 3, 1, 1, 1, 3, 1, 1, 1, 2, 1, 1, 1, 1, 2, 3, 1, 2, 1, 3, 1, 3, 3, 1, 1
PLAN #4 | 1, 1, 1, 1, 1, 1, 1, 1, 1, 1, 3, 1, 1, 1, 3, 1, 1, 1, 2, 1, 1, 1, 1, 2, 3, 1, 2, 1, 1, 1, 3, 3, 1, 3
  ----------------------------------------------------------------------------------------
----- Each line shows a release plan and each number is the release that a feature should be offered in regarding this plan -----
>>
```

As the result of running SDO for two releases in our case study, we came up with four release plans. These four plans are Pareto solutions, and there is no other plan that is better in one criterion and not worse in another criterion.